\documentclass{article}
\usepackage{amsmath}
\usepackage[pdftex]{graphicx}
\usepackage{amsfonts}
\usepackage{amssymb}
\usepackage{ulem}

\begin{document}

\title{Paired states of interacting electrons in a two dimensional lattice}
\author{D.Souza and F.Claro \\
{\small Facultad de F\'isica}\\
{\small Pontificia Universidad Cat\'olica de Chile}\\
{\small Casilla 306, Santiago 22, Chile}\\
[-0.25in]}
\date{}
\maketitle

\begin{abstract}
\noindent We show that two tight binding electrons that repel may form a
bounded pair in two dimensions. The paired states form a band with energies
that scale like the strength of the interaction potential. By applying an
electric field we show that the dynamics of such states is that of a
composite particle of charge 2e. The system still sustains Bloch-like
states, so that if the two bands overlap single and paired states might
coexist allowing for a bosonic fluid component that, if condensed, would
decrease the resistance at low temperatures. The presence of two bands
allows for new oscillations whose experimental detection would permit a
direct measurement of the interaction potential strength.
\end{abstract}

\section*{I. Introduction}

Bound states are normally associated with basins of attraction, and it is
always surprising to find them in the presence of a repulsive interaction.
The most remarkable case is the pairing of electrons in normal
superconductors, mediated by lattice distortions. A new kind was recently
observed for ultracold rubidium atoms in an optical lattice [1], where the
binding arises from pure quantum interference. The pairs were shown to be
stable, thus suggesting that they might form a superfluid phase if the
composite is a boson.

The effect was originally predicted for a one dimensional string [2-5]. The
eigenstates decay exponentially in the relative coordinate, while in the
center of mass coordinate they are extended over the whole lattice. In
higher dimensions one can argue that paired states may exist as well. To see
this, consider two particles that repel moving in a two dimensional lattice.
Together, the pair combine four spatial degrees of freedom, thus allowing
that it be formally described as a single particle moving in a 4D lattice. A
particle-particle interaction appears then as an interface potential in four
dimensions, which decays to both sides of the hyperplane x$_{1}$= x$_{2}$, y$%
_{1}$= y$_{2}$, where (x$_{1}$,y$_{1}$), (x$_{2}$,y$_{2}$) are cartesian
coordinates for the particles position in real 2D space. Surface states
bounded to the interface are then expected even if the potential locally
rises, as would occur under particle-particle repulsion. Since the interface
represents matching coordinates for the two particles in 2D such surface
states are paired states in the lattice. A similar argument applies in 3D.

We find the 2D case particularly interesting since the onset of pairing
could be relevant to high temperature superconductors where transport is
related to conducting sheets, and to certain 2D experimental probes for
which the physics is still controversial [6-16]. It is believed that in the
latter case localization due to disorder on the conducting sheet should
produce a divergent resistivity as the temperature is lowered, but this
behavior was not observed at all electron densities in
metal-oxide-semiconductor field-effect transistors (MOSFET). In fact, a
metal insulator transition was found in low density 2D samples as the
density is slightly increased [8]. An early prediction invoked the strongly
interacting nature of a low density electron fluid [9-10], while more recent
theories study the effect of formation of a Wigner solid [13] or a
superconducting phase [14], a strong spin-orbit interaction [15] and a few
classical effects [16], none of which has received consensus for describing
the physics at the root of the transition [17].

As we shall show, paired singlet states are predicted to exist in 2D for
repelling particles in a lattice. In Section II we define the model and show
that the density of states reveals the presence of two bands of different
character, one corresponding to a Bloch particle in 4D and one describing a
2D surface state. In Section III we show that the dynamics of such surface
state is that of a composite boson of charge 2e, our main finding, and in
Section IV we discuss our results.

\section*{II. The Model}

Consider for definiteness two interacting tight-binding electrons in 2D in a
uniform external field. {The asociated Hamilton operator for such system
reads:} 
\begin{equation}
H=\lambda \sum_{x,y}\sum_{s}(c_{x,y+1,s}^{\dagger
}c_{x,y,s}+c_{x,y,s}^{\dagger }c_{x,y+1,s}+c_{x+1,y,s}^{\dagger
}c_{x,y,s}+c_{x,y,s}^{\dagger }c_{x+1,y,s})  \notag
\end{equation}%
\begin{equation}
+eEa\frac{\widehat{n}\cdot \overrightarrow{R}}{\sqrt{2}}\sum_{x,y}%
\sum_{s}c_{x,y,s}^{\dagger }c_{x,y,s}+U\sum_{x,y}c_{x,y,\uparrow }^{\dagger
}c_{x,y,\uparrow }c_{x,y,\downarrow }^{\dagger }c_{x,y,\downarrow },
\end{equation}%
where $c_{x,y,s}^{\dagger }$ and $c_{x,y,s}$ are the creation and
annihilation operators for one electron located at the site $\left(
x,y\right) $ with spin $s=\uparrow ,\downarrow $, $\lambda $ the usual
hopping energy parameter, $\overrightarrow{R}$ the sum of position vectors, $%
e$ the charge of the electron, $E$ and $\widehat{n}$ the electric field
magnitude and direction, respectively, and $a$ the lattice parameter. The
last term represents a two-body Hubbard contact interaction potential $U$.
We assume that the interaction is strongly screened and may be ignored for
particle-particle separations beyond a lattice constant and discuss only the
singlet state [3].

In the Wannier representation, the time dependent wave function for the pair
in a singlet state may be expanded as%
\begin{equation}
\left\vert \phi \left( t\right) \right\rangle
=\sum_{x_{1},y_{1}}\sum_{x_{2},y_{2}}f_{x_{1},y_{1};x_{2},y_{2}}\left(
t\right) \left\vert x_{1},y_{1},s_{1};x_{2},y_{2},s_{2}\right\rangle 
\end{equation}%
where the sum runs over all lattice sites, $f_{x_{1},y_{1};x_{2},y_{2}}%
\left( t\right) $ is the amplitude for an electron with spin $s_{1}$ to be
at ($x_{1},y_{1}$), and another with spin $s_{2}\neq $ $s_{1}$ to be at ($%
x_{2},y_{2}$), while the ket $\left\vert
x_{1},y_{1},s_{1};x_{2},y_{2},s_{2}\right\rangle $ represents such state.
The amplitude obeys the Schr\"{o}dinger equation of motion%
\begin{equation}
i\hbar \frac{d}{dt}f_{x_{1},y_{1};x_{2},y_{2}}=-\lambda \tilde{f}%
_{x_{1},y_{1};x_{2},y_{2}}+\left( eEaX+U\delta _{x_{1},x_{2}}\delta
_{y_{1},y_{2}}\right) f_{x_{1},y_{1};x_{2},y_{2}},  \label{eq_mov}
\end{equation}%
where $\tilde{f}$ \ is the sum of amplitudes of all nearest neighbors to the
site ($x_{1},y_{1},x_{2},y_{2})$ in the lattice, and $%
X=(x_{1}+y_{1}+x_{2}+y_{2})/\sqrt{2}$ in units of the lattice constant. We
have assumed that the electric field is along the diagonal $x_{1}=y_{1},$ $%
x_{2}=y_{2}$. 

By making the replacement

\begin{equation*}
f_{x_{1},y_{1};x_{2},y_{2}}=e^{ik_{x}(x_{1}+x_{2})}e^{ik_{y}(y_{1}+y_{2})}g(x_{2}-x_{1},y_{2}-y_{1})
\end{equation*}%
the associated eigenvalue equation in the absence of an external field
becomes%
\begin{equation}
Eg(u,v)=-2\lambda \left[ \cos k_{x}a\left( g(u-1,v\right) +g(u+1,v)) +
\cos k_{y}a\left( g(u,v-1\right) +g(u,v+1))\right] 
\notag
\end{equation}
\begin{equation}
+\delta _{u,0}\delta_{v,0}Ug(u,v)
\label{eigen}
\end{equation}
where now $u=x_{2}-x_{1},v=y_{2}-y_{1}.$One can easily check that $%
g(u,v)=\delta _{u,0}\delta _{v,0}$ is an eigenvalue of this equation at the
lowest band edge $k_{x}=0=k_{y}$, of eigenvalue $E=U$. This solution
represents extreme pairing, when one particle is precisely on top of the
other.

In the absence of interaction solutions of Eq. (4) are plane waves in a
Bloch band, with eigenvalues $E=-4\lambda (\cos k_{x}a+\cos k_{y}a),$ where $%
\overrightarrow{k}=(k_{x},k_{y})$ is the center of mass wave number. The
band remains when the repulsive interaction is turned on, while additional
states appear at positive energy. This is seen in Fig. 1, where the density
of states obtained by numerical diagonalization of the eigenvalue equation
for several positive values of the parameter U is shown. Data corresponds to
a finite lattice with N=49 plaquettes using periodic boundary conditions. 
{The bell shaped curve represents the density of states of a
single particle in 4D, the analog problem in this case, while a 2D-like
density of states profile representing surface states in 4D moves away as U
grows. Note that the lower edge of this band is at }${E=U}$%
{\ as expected.}

\begin{figure}[h]
\begin{center}
\includegraphics[scale=0.3]{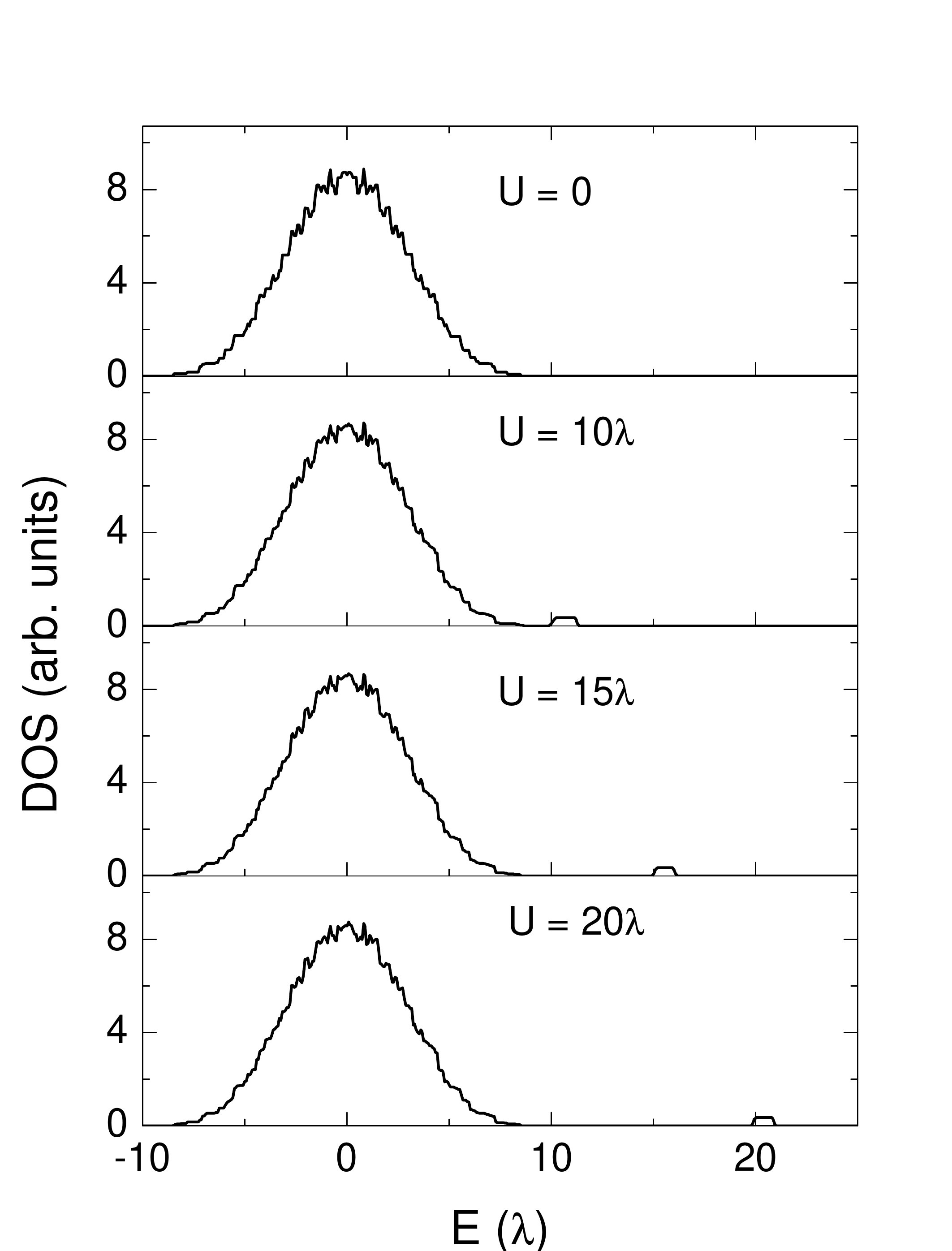}
\end{center}
\caption{Density of states of two interacting electrons in a 2D lattice of
49 sites. For $U=0$, it is that of a single tight-binding particle in 4D.
When $U\neq 0$ a new subband emerges, moving to higher energies as $U$ is
increased. }
\label{fig-dos}
\end{figure}

\section*{III. Evidence for bounded pairs}

To confirm the presence of two different kinds of states, we have calculated
the average distance between particles%
\begin{equation}
d_{j}=a\sum_{x_{1},x_{2}}\sum_{y_{1},y_{2}}\sqrt{\left( x_{1}-x_{2}\right)
^{2}+\left( y_{1}-y_{2}\right) ^{2}}\left\vert
f_{x_{1},y_{1};x_{2},y_{2}}^{j}\right\vert ^{2}.  \label{ec-dj}
\end{equation}%
on a finite 2D sample of N plaquetes using periodic boundary conditions.
Here $j$ labels the eigenstate considered. The results are shown in Fig. 2,
where the quantity $d_{j}$ is shown as a function of sample size both for
states in the B-band as well as states in the U-band. We see that the
average separation of the particles is essentially null for states in the
U-band for all cell sizes, as is expected for a bounded pair. For states in
the Bloch band the separation is finite, with a value that grows linearly
with $N^{\frac{1}{2}}$ as expected for extended states in two dimensions.

\begin{figure}[h]
\par
\begin{center}
\includegraphics[scale=1.0]{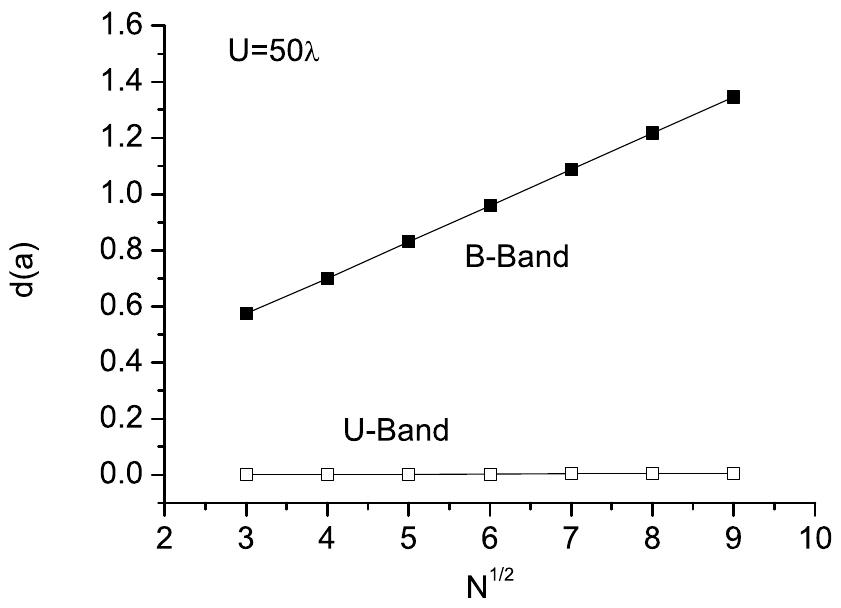}
\end{center}
\caption{Average mean distance of\ Bloch and U states plotted for different
lattice sizes and a contact potential $U=50\protect\lambda $. The mean
separation in a Bloch state (B-band) grows linearly with the edge size,
while for the U-band it stays near zero as expected for paired states.}
\label{fig-prom}
\end{figure}
\qquad \qquad\ \ \ \ \ \ 

A more dramatic confirmation that repelling electrons in a lattice form
bounded pairs is found when the electric field is turned on and the time
evolution of a given initial state is followed by solving Eq. (3). It is
well known that charged particles in a lattice experience Bloch oscillations
in the presence of the field, with a frequency proportional to the particle
charge[18]. To find out if such oscillations are present we studied the time
evolution of the average position along the x axis of each particle%
\begin{equation}
R_{i}^{x}\left( t\right) =a\sum_{x_{1},y_{1}}\sum_{x_{2},y_{2}}\left\vert
f_{x_{1},y_{1};x_{2},y_{2}}(t)\right\vert ^{2}x_{i},  \label{ec-R}
\end{equation}%
where $i=1,2,$ is the particle index, and a similar equation for the
coordinate $y.$ The numerical work was done using a half implicit numerical
method which is second-order accurate and unitary [19]. The positions were
indeed found to oscillate, not simply performing Bloch-type oscillations but
a more complex pattern depending upon initial conditions. Figure \ref%
{fig-fourier_3} shows the power spectrum for a 2D lattice of $N=81$ sites
and an electric field energy $F=eEa=10\lambda $, set rather large in order
to avoid reflections from the edges in our reduced numerical sample. In the
absence of interactions only the Bloch frequency $\omega =F$ ($\hbar =1$)
has any weight, as seen in Fig. 3(a). We next set the interaction strength
at $U=100\lambda ,$ a large value appropriate to have a U-band well
separated from the extended states. For an initial condition with zero
amplitude save for points far from the interaction region $x_{1}=x_{2}$, $%
y_{1}=y_{2}$, the situation is unchanged. But for an initial condition with
unit amplitude in sites over the interaction hyperplane in the 4D space of
coordinates $(x_{1},y_{1},x_{2},y_{2})=(r,s,r,s)$, $r,s=4,5,6,$ and either $%
r $ or $s=5$ when $r\neq s,$ representing a portion of the hyperplane and a
few nearby points, we see a Bloch oscillation with twice the Bloch
frequency, with a power spectrum as shown in Fig. 3(b). We interpret this
new frequency as arising from motion of a composite particle of charge $2e$
in an electric field, thus confirming that the pair behaves dynamically as a
stable composite in the absence of dissipation. Additional
interaction-induced oscillations (ININO) appear near $U+F$ and $U-F,$ shown
in more detail in the inset. These latter oscillations, seen also in 1D
systems, are the result of pure correlated electron transport [20]. We note
that our numerical results show that the features just described persist as
the interaction strength enters the region $U\sim \lambda ,$ only that
hybridization of localized and extended states makes the power spectrum more
complex. 
\begin{figure}[tbp]
\par
\begin{center}
\includegraphics[scale=1.0]{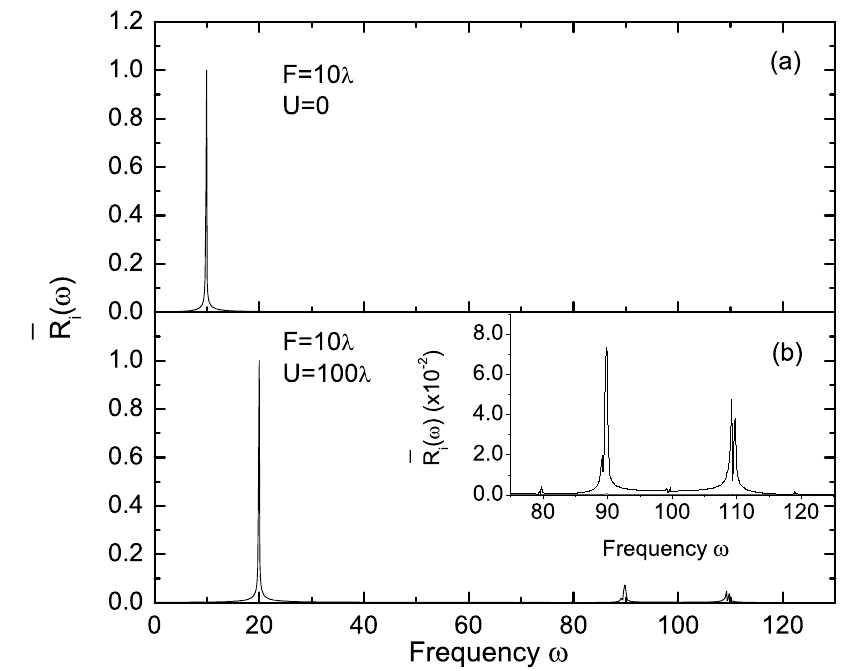} 
\end{center}
\caption{Frequency spectrum of the average position of each particle in a 2D
lattice of 81 sites and an external field of strength $F=10\protect\lambda $%
. (a) is in the absence of interactions and (b) is for a Hubbard interaction
strength $U=100\protect\lambda $. The inset shows details near the Hubbard
frequency. Frequencies are in units of $\protect\lambda $.}
\label{fig-fourier_3}
\end{figure}

\section*{IV. Discussion}

In summary, we have shown that two interacting electrons moving in a lattice
in 2D have singlet paired states grouped in a band, with the dynamics of a
composite particle of charge 2e. Our results were obtained for a highly
screened contact repulsive potential, yet studies in 1D show that a Coulomb
tail does not destroy the pairing [2] in the singlet state. Although the
specific treatment refers to electrons in a lattice, the main results may be
extended to other systems such as atoms in an optical lattice [1].

In order to understand the origin of the paired states we recall that
ordinary Bloch states may be thought of as atomic states that hybridize with
lattice neighbors, forming a tunneling network capable of sustaining
extended states that are grouped in bands. Similarly, the paired eigenstates
may be described as a high energy two-electron single well state that forms
a band of extended states owing to pair tunneling throughout the lattice.
The Hubbard model captures the escence of this picture, but details such as
possible molecular two-electron extended states may arise if a finite range
interaction is included.

The energy of the paired states scales like the interaction strength $U$. If
this latter quantity is of the order of the bandwidth, such states may lie
partly or wholly within the band of extended states, allowing pairs to form
in the ground state if there are enough electrons in the sample. The paired
states equal in number the extended states, since they correspond to single
particle surface states over a 2D planar interface in a 4D lattice. We thus
speculate that if condensation of such pairs indeed occurs then a transition
to low resistance should take place as the density is increased, as is
actually observed in experiment [11].

Our model yields a pair whose dynamics is that of a stable composite of
charge 2e. Such stability may be reduced by dissipation mechanisms such as
phonon excitation. In a static periodic potential such as that generated by
an optical lattice phonons are not present, and a pair of energy $\sim $ $%
U>8\lambda $ above the band center is expected to be stable since its decay
into a band state is forbiden by energy conservation. Under an external
field $F$, transitions between extended and localized states give rise to
oscillations. If the associated frequencies $\omega _{+}=U+F$ and $\omega
_{-}=U-F$ were accessible to experiment, then the sum $\omega _{+}+$ $\omega
_{-}=2U$ would yield a direct measure of the interaction strength in the 2D
sample.

This research was supported by Fondecyt Grant 1060650.

[1] K. Winkler, G. Thalhammer, F. Lang, R. Grimm, J. Hecker Denschlag, A.J.
Daley, A. Kantian, H.P. B\"{u}chler, and P. Zoller, Nature 441, 853 (2006)

{[2] F. Claro, J. F. Weisz and S. Curilef, Phys. Rev. B67, 193101 (2003) }

{[3] J. Weisz and F. Claro, J. Phys.:Condens.Matter 15, 1 (2003) }

{[4] L. Jin, B. Chen, and Z. Song, Phys. Rev. A 79, 032108 (2009) }

{[5] S. M. Mahajan and A. Thyagaraja, J. Phys. A 39, L667 (2006) }

{[6] E. Abrahams, P. W. Anderson, D. C. Licciardello, and T. V.
Ramakrishnan, Phys. Rev. Lett. 42, 673 (1979). }

[7] {B. L. Altshuler, A. G. Aronov, and P. A. Lee, Phys. Rev. Lett. 44, 1288
(1980). }

{[8] S. V. Kravchenko, G. V. Kravchenko, J. E. Furneaux, V. M. Pudalov, M.
DIorio, Phys. Rev. B 50, 8039 (1994); S.V. Kravchenko, W.E. Mason, G.E.
Bowker, J.E. Furneaux, V.M. Pudalov, M. DIorio, Phys. Rev. B 51, 7038
(1995); S.V. Kravchenko, D. Simonian, M.P. Sarachik, W. Mason, J.E.
Furneaux, Phys. Rev. Lett. 77, 4938 (1996). }

{[9] A. M. Finkelshtein, Z. Phys. B 56, 189 (1984); C. Castellani, C. Di
Castro, P. A. Lee, and M. Ma, Phys. Rev. B 30, 527 (1984); C. Castellani, C.
D. DiCastro, and P. A. Lee, Phys. Rev. B 57, R9381 (1998).}

{[10] Q. Si and C. M. Varma, Phys. Rev. Lett. 81, 4951 (1998). }

{[11] V. Dobrosavljevic, E. Abrahams, E. Miranda, and S. Chakravarty, Phys.
Rev. Lett. 79, 455 (1997). }

{[12] S. Chakravarty, L. Yin, and E. Abrahams, Phys. Rev. B 58, R559 (1998). 
}

{[13] S. Chakravarty, S. Kivelson, C. Nayak, and K. Voelker, Phil. Mag. B79,
859-68 (1999) }

{[14] P. Phillips et al., Nature 395, 253 (1998); D. Belitz and T. R.
Kirkpatrick, Phys. Rev. B 58, 8214 (1998); J. S. Thakur and D. Neilson,
Phys. Rev. B 58, 13 717 (1998). }

{[15] Y. Lyanda-Geller, Phys. Rev. Lett. 80, 4273 (1998). }

{[16] S. He and X. C. Xie, Phys. Rev. Lett. 80, 3324 (1998); B. L. Altshuler
and D. L. Maslov, Phys. Rev. Lett. 82, 145 (1999); T. M. Klapwijk and S. Das
Sarma, preprint cond-mat/9810349 (1998) (Solid State Commun., in press); S.
Das Sarma and E. H. Hwang, Phys. Rev. Lett. 83, 164-7 (1999). }

{[17] For a review see S.V. Kravchenko and M.P. Sarachik, Rep. Prog. Phys.
67, 1 (2004) .}

[18] E.E. Mendez, F. Agull\'{o}-Rueda, J.M. Hong, Phys. Rev. Lett. 60, 2426
(1988).

{[19] Abraham Goldberg, Harry M. Schey and Judah L. Schwartz, Am. J. Phys.
35, 177 (1967). }

{[20] See Ref. 2, Eq. 6. A Fourier expansion of this expression yields
oscillations at frequencies U+F and U-F. }

\end{document}